\documentclass[a4,11pt]{article}
\usepackage{epsfig}

\usepackage{amssymb,epsfig,graphicx,epsf}

\usepackage{amsfonts}
\usepackage{amsmath}
\usepackage{amssymb}
\usepackage{mathrsfs}
\usepackage{tensor}
\usepackage{slashed}
\usepackage{enumerate}
\newcommand{\be}{\begin{equation}}
\newcommand{\ee}{\end{equation}}

\newcommand{\ba}{\begin{eqnarray}}
\newcommand{\ea}{\end{eqnarray}}

\begin{document}

\title{Bosonic $(p-1)$-forms in Einstein-Cartan theory of gravity}

\author{Jorge Alfaro $^{(a)}$\thanks{jalfaro@uc.cl} and Simon Riquelme M.$^{(a)(b)}$\thanks{sdriquel@umd.edu}\\
\\
$^{(a)}$Facultad de F\'isica,\\
Pontificia Universidad Cat\'olica de Chile, Casilla 306, Santiago 22, Chile,\\
\\
$^{(b)}$Center for Fundamental Physics, Department of Physics,\\
University of Maryland, College Park, MD 20740-4111, USA}

\date{}

\maketitle

\begin{abstract}
We introduce bosonic $(p-1)$-form fields that couple to the spin connection of the Einstein-Cartan theory of gravity thus becoming a non-trivial source of space-time torsion. We analyze all the general features of both the matter and the gravitational sectors of the theory. Finally we briefly consider the implications of the existence of such fields in different physical settings.

\end{abstract}

\newpage

\section{Introduction}
General Relativity (GR) is a pseudo-Riemannian metric theory of gravity. As such it has been very successful in explaining an impressive amount of astrophysical data. Until now it is the best theory of gravity in the sense of accuracy and economy that we have at our disposal \cite{will}. However, from the theoretical perspective it is an incomplete theory. We are not addressing the well-known fact that GR is a non-renormalizable theory \cite{thooft}, even if it can be taken as an effective field theory \cite{donog}. These are all `quantum' considerations. Even at the strict classical level there is a theoretical clash. The basic assumption of GR is that geometry of the spacetime manifold is of the Riemannian type with the metric $\tensor{g}{_\mu_\nu}$ as the fundamental ingredient for its dynamical description. However, when dealing with fermions in curved spacetime it is unavoidable the introduction of a vielbein field $\tensor*{e}{^a_\mu}$ such that $\tensor{g}{_\mu_\nu} = \tensor*{e}{^a_\mu}\tensor*{e}{^b_\nu}\tensor{\eta}{_a_b}$ where $\tensor{\eta}{_a_b}$ is an `internal' Minkowski metric. The internal Lorentz symmetry of the vielbein in the $a$ index is a local one so it is a gauge symmetry. As such there is an associated gauge field $\tensor*{\omega}{^a^b_\mu}$ which accounts for independent degrees of freedom from those of the metric.\footnote{There is a very nice review of all these ideas in \cite{zanelli}. See also \cite{blagolib} and \cite{ortin}.} Therefore in the spirit of generality fermionic matter is pushing us to consider a Riemann-Cartan geometry instead where spacetime torsion is not set to zero a priori. Since symmetry has historically shown itself as an appealing and usually fundamental principle of nature, we cannot feel comfortable with the fact that only fermions are matter sources of spacetime torsion. Of course Nature will always have the final word on these issues but since spacetime torsion has not yet been experimentally measured in our physical world\footnote{Torsional effects are expected to be significant at the GUT energy scale which is $\sim 10^{16}\,\text{GeV}$ \cite{trautman}.} torsional effects remain as an attractive theoretical idea which can in principle be extended to the bosonic matter sector of field theories which attempt to describe physics.
\\
\\
	In this paper we explore the idea of having bosonic matter that couples to the spin connection of gravity, thus becoming a source for spacetime torsion. In section II we introduce the action principle that describes this whole family of $(p-1)$-forms and their dynamics. Then we explore the conserved currents associated with arbitrary variations of the gravitational fields on this matter sector, i.e., we study the symmetries of the action and the associated conservation laws preserved by the dynamics. Subsequently we consider the geometrodynamical aspects of our theory. Finally in section III we introduce toy models to explore the features that these mathematical objects have in physical theories as in inflation theory and 3-D gravity theory. To conclude we summarize our work and consider ideas for future research.

\section{The Theory}

\subsection{The Action and its Symmetries}
Let us consider the following action:
\begin{align}
S[e,\omega,\phi,p,q] = -\frac{\lambda}{2}\int\,\star(\nabla\tensor{\phi}{_{a_1}_\dots_{a_q}})\wedge\nabla\tensor{\phi}{^{a_1}^\dots^{a_q}}\label{action},
\end{align}
where $\tensor{\phi}{_{a_1}_\dots_{a_q}}$ is a $SO(n)$-valued $(p-1)$-form and $\{a_1,\dots,a_q\}$ is a completely antisymmetric set of indices. As an abstract operator, $\nabla = d + [\omega,\,\,]$ where $d$ is the exterior derivative operator and $\omega$ is the spin gauge connection one-form of gravity \cite{eguchi}. It is evident that $1\leq p\leq n$ and $0\leq q \leq n$ in a $n$-dimensional space. $\lambda$ is a suitable constant. If we define the `field strenght' $p$-form as $\tensor{F}{_{a_1}_\dots_{a_q}} = \nabla\tensor{\phi}{_{a_1}_\dots_{a_q}}$, our action reads
\begin{align}
S[e,\omega,\phi,p,q] &= -\frac{\lambda}{2}\int\,\star\tensor{F}{_{a_1}_\dots_{a_q}}\wedge\tensor{F}{^{a_1}^\dots^{a_q}}\notag\\
&= -\frac{\lambda}{2p!p!}\int\,\tensor{F}{_{a_1}_\dots_{a_q}_{b_1}_\dots_{b_p}}\tensor{F}{^{a_1}^\dots^{a_q}_{c_1}_\dots_{c_p}}\star(e^{b_1}\wedge\dots\wedge e^{b_p})\wedge e^{c_1}\wedge\dots\wedge e^{c_p}\notag\\
&= -\frac{\lambda}{2p!p!(n-p)!}\int\,\tensor{F}{_{a_1}_\dots_{a_q}_{b_1}_\dots_{b_p}}\tensor{F}{^{a_1}^\dots^{a_q}_{c_1}_\dots_{c_p}}\tensor{\epsilon}{^{b_1}^\dots^{b_p}_{b_{p+1}}_\dots_{b_n}}e^{b_{p+1}}\wedge\dots\wedge e^{b_n}\wedge e^{c_1}\wedge\dots\wedge e^{c_p},\notag
\end{align}
and $e^a$ is the vielbein one-form $e^a = \tensor*{e}{^a_\mu}dx^\mu$ such that $\tensor*{e}{^a_\mu}\tensor*{e}{^b_\nu}\tensor{\eta}{_a_b} = \tensor{g}{_\mu_\nu}$, where $\eta$ is the $SO(n)$ inner metric and $g$ is the (possibly curved) space metric.
\noindent If we consider an Einstein-Cartan gravitational theory we get the following currents from this matter theory \cite{gockeler}:
\begin{enumerate}
\item{Energy-momentum}
\begin{align}
\star U_i[e,\omega,\phi,p,q] &= \frac{\lambda}{2(p-1)!}\tensor{F}{_{a_1}_\dots_{a_q}_{b_1}_\dots_{b_{p-1}}_i}\star\tensor{F}{^{a_1}^\dots^{a_q}}\wedge e^{b_1}\wedge\dots\wedge e^{b_{p-1}}\notag\\
&- \frac{\lambda}{2p!(n-p-1)!}\tensor{F}{_{a_1}_\dots_{a_q}_{b_1}_\dots_{b_p}}\tensor{\epsilon}{^{b_1}^\dots^{b_p}_{b_{p+1}}_\dots_{b_{n-1}}_i}\tensor{F}{^{a_1}^\dots^{a_q}}\wedge e^{b_{p+1}}\wedge\dots\wedge e^{b_{n-1}}.\notag
\end{align}
We can write this expression in a more compact way by means of the contraction operator $I$ \footnote{Upon acting on a differential $p$-form $\alpha = \frac{1}{p!}\tensor{\alpha}{_{\mu_1}_{\mu_2}_\dots_{\mu_p}}dx^{\mu_1}\wedge\dots\wedge dx^{\mu_p}$ the contraction operator $I_\xi$ with respect to a vector field $\xi$ is defined by $I_\xi\alpha = \frac{1}{(p-1)!}\xi^{\mu_1}\tensor{\alpha}{_{\mu_1}_{\mu_2}_\dots_{\mu_p}}dx^{\mu_2}\wedge\dots\wedge dx^{\mu_p}$.} \cite{nakahara}. Recalling that $I_{e_i}e^j = \tensor{\delta}{^j_i}$, we get that
\begin{align}
\star U_{i}[e,\omega,\phi,p,q] = \left(\frac{(-1)^{p-1}\lambda}{2}\right)\star\tensor{F}{^{a_1}^\dots^{a_q}}\wedge I_{e_i}\tensor{F}{_{a_1}_\dots_{a_q}} + \left(\frac{(-1)^{n-p}\lambda}{2}\right)\tensor{F}{^{a_1}^\dots^{a_q}}\wedge I_{e_i}\star\tensor{F}{_{a_1}_\dots_{a_q}}.
\end{align}
What we usually call the energy-momentum tensor is defined as $\tensor{\mathcal{T}}{^i_k} = \star(e^i\wedge\star U_k)$.
\item{Spin}
\begin{align}
\star\tensor{J}{_m_n}[e,\omega,\phi,p,q] = (-1)^p\lambda q\star\tensor{F}{_{a_1}_\dots_{a_{q-1}}_m}\wedge\tensor{\phi}{^{a_1}^\dots^{a_{q-1}}_n}.\label{jotastar}
\end{align}
The better known spin tensor is defined as $\tensor{\mathcal{S}}{^i_j_k} = \star(e^i\wedge\star\tensor{J}{_j_k})$.\\
\end{enumerate}
Let us also recall that having a Riemann-Cartan background geometry we can consider the symmetries of a matter theory to get on-shell conservation laws in the sense that \cite{kibble} \footnote{In some references \cite{nieh}\cite{koste} a $\tfrac{1}{2}$ factor enters the definition of the spin current.}
\begin{align}
\delta_\text{symm} S[e,\omega,\phi,p,q] = \int\,\big\{\star U_a\wedge\delta e^a + \star\tensor{J}{_a_b}\wedge\delta\tensor{\omega}{^a^b}\big\} = 0.\label{symmeq}
\end{align}
The coefficient of $\delta\tensor{\phi}{^{a_1}^\dots^{a_q}}$ is zero using \eqref{eom}. The symmetries/conservation laws that we care about are:
\begin{enumerate}[I]
\item{Euclidean symmetry}\\
Under $\delta_E e_a = \delta\tensor{\varepsilon}{_a^b}e_b$ and $\delta_E\tensor{\omega}{^a^b} = -\nabla\delta\tensor{\varepsilon}{^a^b}$ we get
\begin{align}
\nabla\star\tensor{J}{_a_b} + (-1)^{n-1}\star U_{[a}\wedge e_{b]} = 0.\label{eqofcont}
\end{align}
In Einstein-Cartan theory local Euclidean symmetry does not imply a vanishing antisymmetric piece of the energy-momentum tensor $\tensor{\mathcal{T}}{_{[\mu\nu]}}$. Instead it is proportional to the divergence of the spin tensor, $\nabla_\lambda\tensor{\mathcal{S}}{^\lambda_\mu_\nu}\propto\tensor{\mathcal{T}}{_{[\mu\nu]}}$.
\item{Diffeomorphism symmetry}\\
Under $\delta_\text{diff}\,e^a = -\pounds_\xi e^a$ and $\delta_\text{diff}\,\tensor{\omega}{^a^b} = -\pounds_\xi\tensor{\omega}{^a^b}$, where $\pounds_\xi$ stands for the Lie derivative, we get
\begin{align}
\nabla(\star U_a)\xi^a + (-1)^n\star U_a\wedge I_\xi T^a + (-1)^n\star\tensor{J}{_a_b}\wedge I_\xi\tensor{R}{^a^b} = 0.\nonumber
\end{align}
In the last expression we cannot isolate immediately the arbitrary vector field $\xi^\chi$. However it can be shown that it reduces to
\begin{align}
\nabla_\alpha\tensor{\mathcal{T}}{^\alpha_\chi} + \tensor{\mathcal{T}}{^\alpha_\beta}\tensor{T}{^\beta_\alpha_\chi} + \tensor{\mathcal{S}}{^\lambda_\alpha_\beta}\tensor{R}{^\alpha^\beta_\lambda_\chi} = 0.
\end{align}
Here $\nabla_\alpha$ stands for the total covariant derivative which includes torsion. $\tensor{T}{^\alpha_\beta_\gamma}$ is the torsion tensor and $\tensor{R}{^\alpha_\beta_\gamma_\delta}$ is the Riemann tensor. Of course when torsion and spin tensors are set to zero we recover the usual Riemannian covariant energy-momentum tensor conservation law.\\
\end{enumerate}
An arbitrary field variation to this family of actions  is given by $\delta_\phi S[e,\omega,\phi,p,q] = (-1)^{n-p}\lambda\int\,\nabla(\star\nabla\tensor{\phi}{_{a_1}_\dots_{a_q}})\wedge\delta\tensor{\phi}{^{a_1}^\dots^{a_q}} + (-1)^{n-p+1}\lambda\int\,d\,(\star\nabla\tensor{\phi}{_{a_1}_\dots_{a_q}}\wedge\delta\tensor{\phi}{^{a_1}^\dots^{a_q}})$. The first term vanishes by the use of the equations of motion \footnote{Notice that the case $p = 2,q = 0$ gives $d\star F = 0$, i.e., Maxwell's equations. Arbitrary $p\geq3$ for $q = 0$ is the theory of $p$-form electrodynamics \cite{pformelectro}. The case $p=1$, $q=0$ is the massless Klein-Gordon equation for a scalar field, $\Box\,\phi = 0$.}
\begin{align}
\nabla\star\tensor{F}{_{a_1}_\dots_{a_q}} = \nabla\star\nabla\tensor{\phi}{_{a_1}_\dots_{a_q}} = 0.\label{eom}
\end{align}
The second term is the usual boundary term associated to Noether's theorem. It is somehow clear that the conserved current is $\star\tensor{J}{_a_b}$ defined in \eqref{jotastar}. The associated symmetry is that of rotations in Euclidean inner space. As always, the conserved charge is the generator of the symmetry. So we see that in analogy to Maxwell and Yang-Mills theories, when matter couples to the connection it acquires a kind of `gravitational charge', in this case related to the spacetime torsion that it creates \cite{kibble}\cite{hehl}. In other words, space torsion produced by matter coupled to curved geometry can be related to the Noether current associated with inner Euclidean symmetry in tangent space. It can be shown that the Noether current associated to this inner symmetry is given by
\begin{equation}
\star J = (-1)^{n}\varepsilon\star\tensor{J}{_a_b}\tensor{\theta}{^a^b},
\end{equation}
where $\tensor{\theta}{^a^b}$ is an arbitrary set of Euclidean parameters of the transformation. Assuming that the space manifold has a topology $R\times\Sigma$, there is a conserved charge $Q = \int_\Sigma\star J$ which is the generator of the symmetry in the sense of Poisson brackets, $\delta_\text{symm}(\,) = \{\,\,,Q\}_\text{PB}$.\\
\\

\subsection{The Geometry}

Recalling Cartan's structure equations
\begin{align}
T^a &= de^a + \tensor{\omega}{^a_b}\wedge e^b \equiv \nabla e^a,\label{struct1}\\
\tensor{R}{^a^b} &= d\tensor{\omega}{^a^b} + \tensor{\omega}{^a_c}\wedge\tensor{\omega}{^c^b},\label{struct2}
\end{align}
we see that the equation defining the torsion $2$-form is a first order differential equation for the vielbein so we can always find an algebraic solution $\tensor{\omega}{^a^b}[e,\Psi] = \tensor{\widetilde{\omega}}{^a^b}[e] + \tensor{C}{^a^b}[e,\Psi]$ where $\tensor{\widetilde{\omega}}{^a^b}$ is the Riemannian Levi-Civita connection, solution of the homogeneous equation $de^a + \tensor{\widetilde{\omega}}{^a_b}\wedge e^b = 0$ and $T^a = \tensor{C}{^a_b}\wedge e^b$. Here $\Psi(x)$ stands for arbitrary matter fields. It is easy to show that in the case when $\Psi = \phi$ the explicit separation of the action \eqref{action} is given by
\begin{align}
S[e,\omega,\phi,p,q] &= -\frac{\lambda}{2}\int\,\star\tensor{\widetilde{F}}{_{a_1}_\dots_{a_q}}\wedge\tensor{\widetilde{F}}{^{a_1}^\dots^{a_q}} + \int\,\star\tensor{J}{_a_b}\wedge\tensor{C}{^a^b}\notag\\
&+ \frac{\lambda q}{2}\int\,\star(\tensor{C}{_{a_q}^f}\wedge\tensor{\phi}{_{a_1}_\dots_{a_{q-1}}_f})\wedge\tensor{C}{^{a_q}_d}\wedge
\tensor{\phi}{^{a_1}^\dots^{a_{q-1}}^d}\notag\\
&+ \frac{\lambda q(q-1)}{2}\int\,\star(\tensor{C}{_{a_{q-1}}^f}\wedge\tensor{\phi}{_{a_1}_\dots_{a_{q-2}}_f_{a_q}})\wedge\tensor{C}{^{a_{q}}_d}\wedge\tensor{\phi}{^{a_1}^\dots^{a_{q-2}}^{a_{q-1}}^d},\label{sepboson}
\end{align}
where the quantities with $\,\widetilde{}\,$ stand for Riemannian ones (i.e., dependent on the torsion-free Levi-Civita spin connection).\footnote{It is interesting to compare this with the analogue usual case for Dirac fermions with action $S_f[e,\omega,\bar{\psi},\psi] = \frac{i}{2}\int\star e_a\wedge(\bar{\psi}\gamma^a\nabla\psi - \overline{\nabla\psi}\gamma^a\psi)$ where $\nabla\psi = d\psi - \frac{i}{4}\tensor{\omega}{^a^b}\tensor{\sigma}{_a_b}\psi$ and $\tensor{\sigma}{_a_b} = \frac{i}{2}[\gamma_a,\gamma_b]$, $\gamma^a$ being Dirac gamma matrices. In such a theory we find that $\star\tensor{J}{_a_b}[e,\bar{\psi},\psi] = \frac{1}{4}\tensor{\epsilon}{_a_b_c_d}\star e^c\tensor*{j}{^d_A}$ where $\tensor*{j}{^d_A} = \bar{\psi}\gamma_5\gamma^d\psi$ and the explicit separation $S_f[e,\omega,\bar{\psi},\psi] = \frac{i}{2}\int\,\star e_a\wedge(\bar{\psi}\gamma^a\widetilde{\nabla}\psi - \overline{\widetilde{\nabla}\psi}\gamma^a\psi) + \int\,\star\tensor{J}{_a_b}\wedge\tensor{C}{^a^b}$. The $\int\star J\wedge C$ term here is a `four Fermi'-like interaction. This feature is quite interesting and has been explored in several contexts \cite{perez}\cite{freedman} because if somehow experimentally measured it would immediately rule out torsionless Einstein gravity. Actually as we will discuss soon enough we can always write $R = \widetilde{R} + \bar{R}$ where $\widetilde{R}$ is the Riemannian curvature and $\bar{R} = \nabla C - C\wedge C$. Thus when we consider a theory of matter coupled to a dynamical spacetime the term $\star J\wedge C$ coming from the matter sector will always cancel with the term proportional to $\nabla C$ coming from the gravitational sector by using the equation of motion for the spin connection $\omega = \widetilde{\omega} + C$. In the case of fermionic matter, the four-fermi-like term then comes from the gravitational sector through the term proportional to $C\wedge C$ while for our bosonic fields the matter sector will also contribute to interaction terms as can be seen clearly from equation \eqref{sepboson}.}\\
The geometrodynamical theory we want to consider is Einstein-Cartan theory in 4D determined by the action
 \begin{align}
S[e,\omega,\Psi] = -\frac{1}{2\kappa}\int\star(e_a\wedge e_b)\wedge\tensor{R}{^a^b}(\omega) + S_\text{matter}[e,\omega,\Psi],
\end{align}
where $\kappa = 8\pi G_\text{newton}$.
If we vary this action with respect to the vielbein, \,$\delta_e S = 0$, we find Einstein's equations,
\begin{align}
\star\tensor{R}{^a^b}\wedge e_b = -\kappa\star U^a,
\end{align}
where $\star U^a$ is the energy-momentum $3$-form of the matter fields defined in \eqref{symmeq}.\\
If we now vary with respect to the spin connection, \,$\delta_\omega S = 0$, we find using Palatini's identity $\delta_\omega\tensor{R}{^a^b} = \nabla\delta\tensor{\omega}{^a^b}$ and integrating by parts that
\begin{align}
\nabla\star(e_a\wedge e_b) = -2\kappa\star\tensor{J}{_a_b},\label{eom2}
\end{align}
where $\star\tensor{J}{_a_b}$ is the spin-torsion $3$-form of the matter fields also defined through \eqref{symmeq}.\\
Using \eqref{struct1} in \eqref{eom2} we get that
\begin{align}
\frac{1}{2}\tensor{\epsilon}{_a_b^c^d}\,\tensor{C}{_c^f}\wedge e_f\wedge e_d = - \kappa\star\tensor{J}{_a_b}.\label{eomw}
\end{align}
We will make heavy use of \eqref{eomw} in the upcoming sections recalling a known fact about equivalent Lagrangian theories at the classical level when so-called `auxiliary fields' are present.\footnote{The classical theorem behind the fact that algebraic equations of motion can be pulled back into the action giving a completely equivalent theory \cite{pons}\cite{henneaux} goes like this:\\
\textit{Let $S(q_i,Q_j)$ be an action depending on two sets of dynamical variables, $q_i$ and $Q_j$. The solutions of the dynamical equations are extrema of the action with respect to both sets of variables. If the dynamical equations $\frac{\delta S}{\delta q_i} = 0$ have a unique solution, $q_i^{(0)}(Q_j)$ for each choice of $Q_j$, then the pull-back $S(q_i(Q_j),Q_j)$ of the action to the set of solutions has the property that its extrema are precisely the extrema of the total action $S(q_i,Q_j)$.}}


\section{Toy models}

\subsection{$n = 4$, $p = 1$, $q = 1$.}
Let us consider a Euclidean valued $0$-form, $\phi^a$. The action for this object is ($\lambda = 1$),
\begin{align}
S[e,\omega,\phi,1,1] = -\frac{1}{2}\int\,\star(\nabla\phi_a)\wedge\nabla\phi^a,
\end{align}
its associated energy-momentum $3$-form is
\begin{align}
\star U_i[e,\omega,\phi,1,1] &= \frac{1}{2}\tensor{F}{_a_i}\star F^a - \frac{1}{4}\tensor{F}{_a_b}\tensor{\epsilon}{^b_f_g_i}F^a\wedge e^f\wedge e^g,
\end{align}
and its associated spin-torsion $3$-form is
\begin{align}
\star\tensor{J}{_a_b}[e,\omega,\phi,1,1] &= -\star F_{[a}\phi_{b]} = -\frac{1}{2}\{\star(\nabla\phi_a)\phi_b - \star(\nabla\phi_b)\phi_a\},
\end{align}
where we have used the definition $F^a = \nabla\phi^a = d\phi^a + \tensor{\omega}{^a_b}\phi^b$.\\
Let us see what this implies in Einstein-Cartan theory. The equation we must solve is \eqref{eomw}. So we have
\begin{align}
\frac{1}{2}\tensor{\epsilon}{_a_b^c^d}\tensor{C}{_c^f}\wedge e_f\wedge e_d &= -\kappa\star\tensor{J}{_a_b} = -\kappa\star\tensor{\widetilde{J}}{_a_b} + \frac{\kappa}{2}\{\star(\tensor{C}{_a^f}\phi_f)\phi_b - \star(\tensor{C}{_b^f}\phi_f)\phi_a\},\label{eqb0}
\end{align}
where we define
\begin{align}
\star\tensor{\widetilde{J}}{_a_b} \equiv -\frac{1}{2}\{\star(\widetilde{\nabla}\phi_a)\phi_b - \star(\widetilde{\nabla}\phi_b)\phi_a\}.
\end{align}
Taking the Hodge dual and recalling that for a $p$-form $\omega_p$ in Euclidean space $\star\star\omega_p = (-1)^{p(n-p)}\omega_p$ (so in particular $\star\star e^i = -e^i$), \eqref{eqb0} can be written in the following way:
\begin{align}
\frac{1}{2}\tensor{\epsilon}{_a_b_c_d}\tensor{C}{_c_l_j}\tensor{\epsilon}{_i_j_l_d} - \frac{\kappa}{2}\{\tensor{C}{_a_l_i}\phi_b\phi_l - \tensor{C}{_b_l_i}\phi_a\phi_l\} - \kappa\tensor{J}{_a_b_i} = 0,
\end{align}
where $(\star\star\tensor{\widetilde{J}}{_a_b})_i e^i \equiv \tensor{J}{_a_b_i}e^i$, $\tensor{C}{_a_b_i} = - \tensor{C}{_b_a_i}$ and $\tensor{J}{_a_b_i} = -\tensor{J}{_b_a_i}$.\\
Now we introduce the following notation: Let $\tensor{A}{_i_j_k_\dots_r_s_t}$ be a generic tensor field. From now on we will call it $\tensor{A}{_i_j_k_\dots_r_s_t}\equiv A(i,j,k,\dots,r,s,t)$. In this manner we will shorten the notation for objects like $\tensor{A}{_i_j_k_\dots_r_s_t}\varphi_i\equiv A(\varphi,j,k,\dots,r,s,t)$ where $\varphi_i\equiv \varphi(i)$ is a generic vector field.\\
Our equation becomes
\begin{align}
&\frac{1}{2}C(\phi,a,i)\phi(b)\kappa - \frac{1}{2}C(\phi,b,i)\phi(a)\kappa + \frac{1}{2}C(a,i,b) - \frac{1}{2}C(a,l,l)d(b,i)\notag\\
&- \frac{1}{2}C(b,i,a) + \frac{1}{2}C(b,l,l)d(a,i) - J(a,b,i)\kappa = 0,
\end{align}
where we also denote $d(a,b)\equiv\tensor{\delta}{_a_b}$. The explicit calculation of the solution (which was found with the aid of \textbf{FORM}\cite{vermaseren1}\cite{vermaseren2}) is:
\begin{align}
C(a,b,i) &= \kappa J(a,b,i) + \kappa J(a,i,b) - \kappa J(b,i,a) + \kappa^2\phi(i)J(a,b,\phi)\notag\\
&+\frac{\kappa^2}{[1-\kappa\phi^2]}\bigg\{\phi(a)J(\phi,b,i) + \phi(a)J(\phi,i,b) - \phi(b)J(\phi,a,i) - \phi(b)J(\phi,i,a)\bigg\}\notag\\
&+ \frac{2\kappa}{[2-\kappa\phi^2]}d(b,i)\bigg\{\kappa J(\phi,a,\phi) - \frac{\kappa[2-\kappa^2\phi^4]}{[1-\kappa\phi^2][2+\kappa\phi^2]}\phi(a)J(\phi,l,l) - [1-\kappa\phi^2]J(a,l,l)\bigg\}\notag\\
&- \frac{2\kappa}{[2-\kappa\phi^2]}d(a,i)\bigg\{\kappa J(\phi,b,\phi) - \frac{\kappa[2-\kappa^2\phi^4]}{[1-\kappa\phi^2][2+\kappa\phi^2]}\phi(b)J(\phi,l,l) - [1-\kappa\phi^2]J(b,l,l)\bigg\}\notag\\
&+ \frac{\kappa^2}{[2-\kappa\phi^2]}\phi(b)\phi(i)\bigg\{\frac{\kappa^2\phi^2}{[1-\kappa\phi^2]}J(\phi,a,\phi) - 2J(a,l,l)\bigg\}\notag\\
&- \frac{\kappa^2}{[2-\kappa\phi^2]}\phi(a)\phi(i)\bigg\{\frac{\kappa^2\phi^2}{[1-\kappa\phi^2]}J(\phi,b,\phi) - 2J(b,l,l)\bigg\}.\label{eqfinc}
\end{align}
The total action is then
\begin{align}
S_\text{total}[e,\widetilde{\omega},\phi] &= -\frac{1}{2\kappa}\int\,\star(e_a\wedge e_b)\wedge\tensor{\widetilde{R}}{^a^b} - \frac{1}{2}\int\,\star\widetilde{F}_a\wedge\widetilde{F}^a + \frac{1}{2\kappa}\int\,\star(e_a\wedge e_b)\wedge\tensor{C}{^a_d}\wedge\tensor{C}{^d^b}\notag\\
&+ \frac{1}{2}\int\,\star\tensor{C}{_a^f}\wedge\tensor{C}{^a_d}\,\phi_f\phi^d.
\end{align}
The first two terms are the usual Riemannian ones. The other two are torsional contributions which depend upon the contorsion one-form and through \eqref{eqfinc} they generate a highly non-trivial (self)-interaction potential for the $\phi^a$ field. The appearance of denominators which depend upon $\kappa\phi^2$ is a charasteristic feature of the problem. So for certain configurations of the field $\phi^a$ the interaction terms can grow enormously within the action or the dynamics, recalling that the equation of motion $\nabla\star\nabla\phi_a = 0$ contains torsional terms through the covariant derivative.\\

However, it is almost an impossible task to solve at least analytically the equation of motion for our field and we do not have extra parameters to play with so we will not be able to control the dynamics for our own convenience if for example, we would like to use this toy model as a viable alternative for current inflation theory. The good thing about our toy model is that the potential is univoquely defined, at the classical level, through the algebraic equation of motion for the spin connection and so would be a falsifiable self-contained proposal instead of an ad-hoc ansatz for the `inflaton' potential. It can be argued that a proper Wick rotation will give us back a faithful Lorentzian expression. Let us now consider the implications of such a model applied to inflation theory. 

\subsection{Euclidean(Lorentz)-valued scalar as the inflaton}

Let us consider the following action in Euclidean spacetime
\begin{align}
S[e,\omega,\phi] &= S_G[e,\omega] + S_M[e,\omega,\phi]\notag\\
&= \frac{1}{2\kappa}\int\,\star(e_a\wedge e_b)\wedge\tensor{R}{^a^b}(\omega) - \frac{1}{2}\int\,\star(\nabla\phi_a)\wedge\nabla\phi^a\notag\\
&= \frac{1}{2\kappa}\int d^4x\,e\,R(\omega) + \frac{1}{2}\int d^4x\,e\,\nabla_\mu\phi_\nu\nabla^\mu\phi^\nu.
\end{align}
The algebraic equation of motion for the spin connection can be solved as before leading to first order in $\kappa$ to
\begin{align}
C(a,b,i) = \kappa J(a,b,i) + \kappa J(a,i,b) - \kappa J(b,i,a) - \kappa J(a,l,l)d(b,i) + \kappa J(b,l,l)d(a,i),\label{clin}
\end{align}
where 
\begin{align}
J(a,b,i) \equiv -\frac{1}{2}\{(\widetilde{\nabla}_i\phi_a)\phi_b - (\widetilde{\nabla}_i\phi_b)\phi_a)\}.
\end{align}
Then pulling back \eqref{clin} we get that $S[e,\omega,\phi]\rightarrow S[e,\widetilde{\omega},\phi]$ where
\begin{align}
S[e,\widetilde{\omega},\phi] &= S_G[e,\widetilde{\omega}] + S_M[e,\widetilde{\omega},\phi] + S_\text{int}[e,\widetilde{\omega},\phi]\notag\\
&= \frac{1}{2\kappa}\int\,\star(e_a\wedge e_b)\wedge\tensor{\widetilde{R}}{^a^b}(\widetilde{\omega}) - \frac{1}{2}\int\,\star\widetilde{F}_a\wedge\widetilde{F}^a - \frac{1}{2\kappa}\int\,\star(e_a\wedge e_b)\wedge\tensor{C}{^a_d}\wedge\tensor{C}{^d^b}\notag\\
&+ \frac{1}{2}\int\,\star\tensor{C}{_a^f}\wedge\tensor{C}{^a_d}\,\phi_f\phi^d,
\end{align}
and in components $S_\text{int}$ reads
\begin{align}
S_\text{int} = \int d^4x\,e\,\left\{\frac{1}{2\kappa}\tensor{C}{^a_d_a}\tensor{C}{^b^d_b} + \frac{1}{2\kappa}\tensor{C}{^a_d_b}\tensor{C}{^d^b_a} - \frac{1}{2}\tensor{C}{_a_b^f}\tensor{C}{^a^d_f}\phi^b\phi_d\right\}.\label{sint}
\end{align}
We observe that in the linear approximation $C^2\sim\mathcal{O}(\kappa^2)$ so the last term of the above integrand will not contribute. Thus the interaction terms within this approximation come solely from the gravitational sector of the theory. Keeping the linear terms in $\kappa$ the expression inside the curly brackets of \eqref{sint} reads
\begin{align}
\mathscr{L}_\text{int} &= \kappa J(f,g,g)J(f,h,h) - \frac{1}{2}\kappa J(f,g,h)J(f,g,h) + \kappa J(f,g,h)J(g,h,f)\notag\\
&= -\frac{1}{2}\kappa\phi^\mu\phi^\nu\widetilde{D}_\mu\phi_\nu\widetilde{D}_\rho\phi^\rho + \frac{1}{2}\kappa\phi^\mu\phi^\nu\widetilde{D}_\mu\phi^\rho\widetilde{D}_\rho\phi_\nu + \frac{1}{4}\kappa\phi^\mu\phi^\nu\widetilde{D}_\rho\phi_\mu\widetilde{D}^\rho\phi_\nu + \frac{1}{4}\kappa\phi^2\widetilde{D}_\mu\phi^\mu\widetilde{D}_\nu\phi^\nu\notag\\
&- \frac{1}{4}\kappa\phi^2\widetilde{D}_\mu\phi_\nu\widetilde{D}^\mu\phi^\nu - \frac{1}{4}\kappa\phi^2\widetilde{D}_\mu\phi_\nu\widetilde{D}^\nu\phi^\mu,
\end{align}
where $\phi^\mu \equiv \tensor*{e}{^\mu_a}\phi^a$ and we have used the fact that $\widetilde{\nabla}_\mu\phi^a = \widetilde{\nabla}_\mu(\tensor*{e}{^a_\nu}\phi^\nu) \equiv \tensor*{e}{^a_\nu}\widetilde{D}_\mu\phi^\nu$ where $\widetilde{D}{_\mu}\phi^\nu = \partial_\mu\phi^\nu + \tensor{\widetilde{\Gamma}}{^\nu_\rho_\mu}\phi^\rho$ and $\tensor{\widetilde{\Gamma}}{^\alpha_\beta_\gamma} = \frac{1}{2}\tensor{g}{^\alpha^\delta}(\tensor{g}{_\delta_\beta_{,\gamma}} + \tensor{g}{_\delta_\gamma_{,\beta}} - \tensor{g}{_\beta_\gamma_{,\delta}})$ are the usual symmetric Christoffel symbols associated with the metric tensor $\tensor{g}{_\mu_\nu}$.\\
We see that already at first order in $\kappa$ we get a rich interacting theory. The detailed analysis of such a theory with a realistic Lorentzian signature will be addressed elsewhere \cite{alfaroriquelme}. However even at zeroth order in $\kappa$ there is a non-trivial theory which we can now address in a Lorentzian space-time.\\
So let us consider the action 
\begin{align}
S = \int\,d^4x\sqrt{-g}\,\left\{\frac{1}{2}\widetilde{D}_\mu\phi_\nu\widetilde{D}^\mu\phi^\nu - V(\Phi^2)\right\},
\end{align}
where $\Phi^2 \equiv \phi_\mu\phi^\mu$. This is the limit where we neglect terms of order $\kappa$ in the above theory. We also added a generic potential for the field. Its associated energy-momentum tensor defined through $\sqrt{-g}\,\tensor{T}{^\alpha^\beta}\delta\tensor{g}{_\alpha_\beta} \equiv - 2\,\delta(\sqrt{-g}\mathscr{L}_\phi) = -\sqrt{-g}\,\tensor{g}{^\alpha^\beta}\delta\tensor{g}{_\alpha_\beta}\mathscr{L}_\phi - 2\sqrt{-g}\,\delta\mathscr{L}_\phi$ reads
\begin{align}
\tensor{T}{_\alpha_\beta} &= \widetilde{D}_\alpha\phi_\mu\widetilde{D}_\beta\phi^\mu + \widetilde{D}_\mu\phi_\alpha\widetilde{D}^\mu\phi_\beta - \widetilde{D}_\mu(\phi_{(\alpha}\widetilde{D}_{\beta)}\phi^\mu) - \widetilde{D}_\mu(\phi_{(\alpha}\widetilde{D}^\mu\phi_{\beta)}) + \widetilde{D}_\mu(\phi^\mu\widetilde{D}_{(\alpha}\phi_{\beta)})\notag\\
&+ 2\frac{\delta V(\Phi^2)}{\delta\tensor{g}{^\alpha^\beta}} - \tensor{g}{_\alpha_\beta}\mathscr{L}_\phi.
\end{align}

The equation of motion reads
\begin{align}
\widetilde{D}_\rho(\sqrt{-g}\,\tensor{g}{^\mu^\rho}\tensor{g}{^\nu^\lambda}\widetilde{D}_\mu\phi_\nu) + \sqrt{-g}\,\frac{\delta V(\Phi^2)}{\delta\phi_\lambda} = \sqrt{-g}\,\tensor{g}{^\mu^\rho}\tensor{g}{^\nu^\lambda}\widetilde{D}_\rho\widetilde{D}_\mu\phi_\nu + \sqrt{-g}\,\frac{\delta V(\Phi^2)}{\delta\phi_\lambda} = 0,\label{eqmotinf}
\end{align}
which can be rewritten as
\begin{align*}
&\tensor{g}{^\mu^\rho}\partial_\rho\partial_\mu\phi^\lambda + \tensor{g}{^\mu^\rho}\partial_\rho(\tensor{\Gamma}{^\lambda_\omega_\mu})\phi^\omega + \tensor{g}{^\mu^\rho}\tensor{\Gamma}{^\lambda_\omega_\mu}\partial_\rho\phi^\omega - \tensor{g}{^\mu^\rho}\tensor{\Gamma}{^\omega_\mu_\rho}\partial_\omega\phi^\lambda\notag\\
&- \tensor{g}{^\mu^\rho}\tensor{\Gamma}{^\omega_\mu_\rho}\tensor{\Gamma}{^\lambda_\chi_\omega}\phi^\chi + \tensor{g}{^\mu^\rho}\tensor{\Gamma}{^\lambda_\omega_\rho}\partial_\mu\phi^\omega + \tensor{g}{^\mu^\rho}\tensor{\Gamma}{^\lambda_\omega_\rho}\tensor{\Gamma}{^\omega_\chi_\mu}\phi^\chi + \frac{\delta V(\Phi^2)}{\delta\phi_\lambda} = 0.
\end{align*}
When $\tensor{g}{_\mu_\nu} = \text{diag}(1,-a^2,-a^2,-a^2)$ the only non-vanishing Christoffel symbols are $\tensor{\Gamma}{^i_0_i} = H$ and $\tensor{\Gamma}{^0_i_i} = a^2H$ (no summation in $i$) where $H = \frac{\dot{a}}{a}$ and $\dot{a} \equiv \frac{da}{dt}$. Then, \eqref{eqmotinf} reads
\begin{align}
\ddot{\phi}^0 + 3H\dot{\phi}^0 - 3H^2\phi^0 + \frac{\partial V(\Phi^2)}{\partial\phi_0}&= 0,\label{eominf}\\
\ddot{\phi}^j + 5H\dot{\phi}^j + (\dot{H} + 3H^2)\phi^j + \frac{\partial V(\Phi^2)}{\partial\phi_j}&= 0,
\end{align}
where we have used the approximation of homogeneity, $\partial_i\phi^\mu = 0$ and $\partial_i\tensor{\Gamma}{^\nu_\rho_\sigma} = 0$.\\
\\
From the energy-momentum tensor we can read off the density $\rho$ and the pressure $P$. These quantities are defined as $\rho \equiv \tensor{T}{^0_0}$ and $P \equiv \frac{1}{3}\sum_i P_i \equiv -\frac{1}{3}\sum_i\tensor{T}{^i_i}$. Explicitly they are given by
\begin{align}
\rho &= \frac{1}{2}\dot{\phi^0}^2 - 3H\phi^0\dot{\phi^0} - \phi^0\ddot{\phi^0} + \frac{9}{2}H^2{\phi^0}^2 - \frac{1}{2}a^2\dot{\phi^i}^2 - a^2H\phi^i\dot{\phi^i} - a^2H^2{\phi^i}^2 + V(\Phi^2),\label{density}\\
P &= \frac{1}{2}\dot{\phi^0}^2 - 2H\phi^0\dot{\phi^0} - \frac{3}{2}H^2{\phi^0}^2 - \dot{H}{\phi^0}^2 - \frac{1}{2}a^2\dot{\phi^i}^2 - 2a^2H\phi^i\dot{\phi^i} - \frac{1}{3}a^2\phi^i\ddot{\phi^i} - \frac{1}{3}a^2H^2{\phi^i}^2\notag\label{pressure}\\
&- V(\Phi^2) + \frac{\partial V(\Phi^2)}{\partial\log a}.
\end{align}
We see through the cosmological equation of state $P = w\rho$ that a non-trivial dependence of the form $w = w(\phi,H)$ is generated even when $V = 0$. 
\\
Since the equations of motion are decoupled, we can in principle set $\phi^j = 0$ keeping only the $\phi^0$ field. Using \eqref{eominf} and \eqref{density} in this limit, dropping the $0$ superscript and recalling Friedmann's equation $H^2 = \frac{\kappa}{3}\rho$ we get that $H^2 = \frac{\kappa}{3}\left(\frac{1}{2}\dot{\phi}^2 + \frac{3}{2}H^2\phi^2 + \frac{\partial V}{\partial\log\phi} + V\right)$. Ignoring the kinetic term as in usual slow-roll inflation theory \cite{kinney}\cite{baumann} and taking $V = \frac{1}{2}m^2\phi^2$ we get that
\begin{align}
H^2 \simeq \frac{\kappa}{3}V_{\text{eff}},\quad V_{\text{eff}} = \frac{3m^2\phi^2}{2\left(1 - \frac{\kappa}{2}\phi^2\right)} \simeq \frac{3}{2}m^2\phi^2 + \frac{3}{4}\kappa m^2\phi^4 + \dots
\end{align}
Now using the usual definition of the slow-roll parameters we can calculate the scalar spectral index $n_S$ and the tensor-to-scalar ratio $r$ with the aid of Mathematica \cite{mathe}. A possible set of outcomes is $n_S \simeq 1.023$ and $r \simeq 0.004$.\footnote{Defining the number of e-folds $N$ through $dN \equiv -Hdt$ with the sign convention such that $N$ is large in the far past and decreases as we go forward in time and as the scale factor $a$ increases, and recalling the definition of the potential slow-roll parameter $\epsilon$ as $\epsilon(\phi) = \frac{1}{2\kappa}\left(\frac{V'}{V}\right)^2$ we can calculate $\epsilon(\phi_e) = 1$ so we can use $N = \sqrt{\frac{\kappa}{2}}\int_{\phi_e}^{\phi_N}\frac{d\hat{\phi}}{\sqrt{\epsilon(\hat{\phi})}}$ to get $\phi = \phi_N$. The CMB constraint is $P_\mathcal{R} \equiv \left(\frac{\kappa^3}{12\pi^2}\right)^{1/2}\left(\frac{V^{3/2}}{V'}\right)\big|_{\phi = \phi_N} \sim 10^{-5}$ which implies that $m^2 \sim -(3.32\times10^{12}\text{GeV})^2$. The tensor-to-scalar ratio is defined as $r \equiv \frac{P_{T}}{P_\mathcal{R}} \simeq 16\epsilon(\phi_{60}) \simeq -8n_T$. Finally, recalling the definition $\eta(\phi) = \frac{1}{\kappa}\left[\frac{V''(\phi)}{V(\phi)}-\frac{1}{2}\left(\frac{V'(\phi)}{V(\phi)}\right)^2\right]$ it can be shown that the scalar spectral index is given by $n_S \simeq 1 - 4\epsilon(\phi_{60}) + 2\eta(\phi_{60})$.} Unfortunately these values are not in accordance with Planck \cite{planck} and BICEP2 \cite{bicep2}\cite{dust} collaborations partial results. Further analysis must be done within these class of models to make them physically realistic \cite{alfaroriquelme}.\\
Let us finally consider the implications that the existence of these fields would have in a theory of 3D gravity with torsion.


\subsection{3D gravity with torsion}

Let us consider Lorentzian 3D gravity with a local Lorentz frame metric of the form $\tensor{\eta}{_i_j} = (+,-,-)$. The normalization of the totally antisymmetric tensor is such that $\tensor{\epsilon}{^0^1^2} = 1$. Since in 3D an antisymmetric tensor is dual to a vector, we make the following definitions: $\tensor{\omega}{^i^j} = -\tensor{\epsilon}{^i^j_k}\omega^k$, $\tensor{R}{^i^j} = -\tensor{\epsilon}{^i^j_k}R^k$. Then, Cartan's structure equations become
\begin{align}
T^i &= de^i + \tensor{\epsilon}{^i_j_k}\omega^j\wedge e^k,\\
R^i &= d\omega^i + \frac{1}{2}\tensor{\epsilon}{^i_j_k}\omega^j\wedge\omega^k.
\end{align}
As before, we can split the spin connection in such a way that $\omega^i = \widetilde{\omega}^i + C^i$, where $\widetilde{\omega}^i$ satisfies the homogeneous first structure equation and $C^i$ is the contorsion one-form such that $T^i = \tensor{\epsilon}{^i_m_n}C^m\wedge e^n$. Finally it is easy to show that
\begin{align}
2R_i = 2\widetilde{R}_i + 2\widetilde{\nabla}C_i + \tensor{\epsilon}{_i_m_n}C^m\wedge C^n,\label{split}
\end{align}
where $\widetilde{R}_i$ is the Riemannian curvature.
We will consider a natural generalization of General Relativity with a cosmological constant, the so-called Mielke-Baekler model \cite{mielke}\cite{baekler}\cite{mardo}, namely,
\begin{align}
S_G[e,\omega] = \int\,2ae^i\wedge R_i - \frac{\Lambda}{3}\tensor{\epsilon}{_i_j_k}e^i\wedge e^j\wedge e^k + \alpha_3L_\text{CS}(\omega) + \alpha_4e^i\wedge T_i,
\end{align}
where $a = \frac{1}{16\pi G}$ and $L_\text{CS}(\omega) = \omega^i\wedge d\omega_i + \frac{1}{3}\tensor{\epsilon}{_i_j_k}\omega^i\wedge\omega^j\wedge\omega^k$ is the Chern-Simons Lagrangian for the Lorentz connection.
The complete action will be $S_T[e,\omega,\Psi] = S_G[e,\omega] + S_M[e,\omega,\Psi]$ where $S_M$ stands for the action of arbitrary matter fields $\Psi(x)$. In our case
\begin{align}
S_M[e,\omega,\Psi] = -\frac{\lambda}{2}\int\,\star F_{a_1\dots a_q}\wedge F^{a_1\dots a_q},
\end{align}
where as before $F_{a_1\dots a_q} = \nabla\phi_{a_1\dots a_q}$ with $\phi$ being a $(p-1)$-form such that $1\leq p\leq 3$, $0\leq q\leq 3$. \footnote{One could also consider `Chern-Simons'-like terms $\sim\phi\underbrace{\nabla\phi\nabla\phi\dots\nabla\phi}_{r\,\,\text{times}}$ which would be only allowed when $n = p(r + 1)-1$ with $r$ some positive integer. For $n = 3$, we would need that $p = 1$ and $r = 3$ or $p = 2$ and $r = 1$ which means $\phi$ should be a 0-form or a 1-form respectively. We could also include terms like $\sim\underbrace{\phi\phi\phi\dots\phi}_{m\,\,\text{times}}$ when $n = m(p-1)$ with $m$ some positive integer (in both cases taking care of the different ways one can hook up the Lorentz indices which is a constraint on $q$) as well as $\sim\star\phi\phi$ which would stand for a `mass term'. Finally one could consider terms like $\sim\nabla\phi\nabla\phi$ when $n = 2p$. Needless to say the terms without the $\star$ do not need the notion of a spacetime metric and thus are `topological' in nature.}\\
The equations of motion are
\begin{align}
2aR_i + 2\alpha_4T_i - \Lambda\tensor{\epsilon}{_i_j_k}e^j\wedge e^k &= \Theta_i,\\
2\alpha_3R_i + 2aT_i + \alpha_4\tensor{\epsilon}{_i_j_k}e^j\wedge e^k &= \Sigma_i,\\
\nabla\star F_{a_{1}\dots a_{q}} = \nabla\star\nabla\phi_{a_{1}\dots a_{q}} &= 0,\label{eqphi}
\end{align}
where $\Theta_i = -\frac{\delta L_M}{\delta e^i}$, $\Sigma_i = -\frac{\delta L_M}{\delta\omega^i}$ are the current $2$-forms due to the presence of the matter field $\phi_a$. Following \cite{blago} when $\Delta \equiv \alpha_3\alpha_4 - a^2 \neq 0$ the first two equations can be rewritten as
\begin{align}
2T_i - \mathfrak{p}\tensor{\epsilon}{_i_j_k}e^j\wedge e^k &= \mathfrak{u}\Theta_i - \mathfrak{v}\Sigma_i,\\
2R_i - \mathfrak{q}\tensor{\epsilon}{_i_j_k}e^j\wedge e^k &= -\mathfrak{v}\Theta_i + \mathfrak{w}\Sigma_i,
\end{align}
where $\mathfrak{p} \equiv \frac{\alpha_3\Lambda + \alpha_4 a}{\Delta}$, $\mathfrak{q} \equiv - \frac{(\alpha_4)^2 + a\Lambda}{\Delta}$, $\mathfrak{u} \equiv \frac{\alpha_3}{\Delta}$, $\mathfrak{v} \equiv \frac{a}{\Delta}$, $\mathfrak{w} \equiv \frac{\alpha_4}{\Delta}$. Remembering that the energy-momentum tensor is $\tensor{\mathcal{T}}{^i_k} \equiv \star(e^i\wedge\Theta_k)$ we can express the energy-momentum $2$-form as
\begin{align}
\Theta_i &= \frac{1}{2}\tensor{\mathcal{T}}{^k_i}\tensor{\epsilon}{_k_m_n}e^m\wedge e^n = \tensor{\epsilon}{_i_m_n}t^m\wedge e^n,\\
t^m &= -\left(\tensor{\mathcal{T}}{^m_k} - \frac{1}{2}\tensor{\delta}{^m_k}\mathcal{T}\right)e^k,
\end{align}
where $\mathcal{T} = \tensor{\mathcal{T}}{^k_k}$.\\
Equivalently since $\tensor{\mathcal{S}}{^k_i} = \star(e^k\wedge\Sigma_i)$, we can write
\begin{align}
\Sigma_i &= \frac{1}{2}\tensor{\mathcal{S}}{^k_i}\tensor{\epsilon}{_k_m_n}e^m\wedge e^n = \tensor{\epsilon}{_i_m_n}s^m\wedge e^n,\\
s^m &= -\left(\tensor{\mathcal{S}}{^m_k} - \frac{1}{2}\tensor{\delta}{^m_k}\mathcal{S}\right)e^k,
\end{align}
where $\mathcal{S} = \tensor{\mathcal{S}}{^k_k}$. Using these results in the equation of motion for $T_i$ we find that
\begin{align}
C^j = \frac{1}{2}(\mathfrak{p}e^j + \mathfrak{u}t^j - \mathfrak{v}s^j).
\end{align}
Using this fact in the second equation of motion we get that
\begin{align}
2R_i = \mathfrak{q}\tensor{\epsilon}{_i_j_k}e^j\wedge e^k - \mathfrak{v}\tensor{\epsilon}{_i_j_k}t^j\wedge e^k + \mathfrak{w}\tensor{\epsilon}{_i_j_k}s^j\wedge e^k.
\end{align}
Recalling the splitting between Riemannian and torsional contributions \eqref{split} we get that
\begin{align}
2R_i =& 2\widetilde{R}_i + \mathfrak{u}\widetilde{\nabla}t_i + \frac{\mathfrak{p}^2}{4}\tensor{\epsilon}{_i_j_k}e^j\wedge e^k + \tensor{\epsilon}{_i_j_k}\left(\frac{\mathfrak{u}\mathfrak{p}}{2}t^j\wedge e^k + \frac{\mathfrak{u}^2}{4}t^j\wedge t^m\right)\notag\\
&- \mathfrak{v}\widetilde{\nabla}s_i - \tensor{\epsilon}{_i_j_k}\left(\frac{\mathfrak{v}\mathfrak{p}}{2}s^j\wedge e^k + \frac{\mathfrak{u}\mathfrak{v}}{2}t^j\wedge s^k - \frac{\mathfrak{v}^2}{4}s^j\wedge s^m\right).\label{riem3d}
\end{align}
In this form of the gravitational field equations, the role of $\phi_i$ as a source of gravity is clearly described by the one-forms $t_i$ and $s_i$. The second line of equation \eqref{riem3d} was not present in \cite{blago}. It is certainly a new feature of our $(p-1)$-forms as torsional bosonic sources.\\
Together with the equations of motion for the matter fields \eqref{eqphi} and a suitable set of boundary conditions define the complete dynamics of the gravitational and matter fields.

\section{Summary}
In this paper we have taken very seriously the fact that Lorentz invariance of physics must be a local (gauge) symmetry as it is the only way to consider fermionic matter living in a curved spacetime background. This observation implies immediately that in principle anything that `feels' the gravitational (spin) connection acquires a `charge' in the same manner that in usual gauge theories only charged matter fields couple with the gauge bosons. This charge is the space integral of the time component of a current that is proportional to the spacetime torsion that the presence of these fields generate \cite{kibble}\cite{hehl}.\\
\\
Accordingly we have built a family of actions for bosonic $(p-1)$-forms that generate spacetime torsion in a very non-trivial way and proved that the previous statement indeed holds in this case as well. We calculated all significant (classical) consequences of having such fields around as conserved quantities, equations of motion and `effective' actions.\\
\\
Finally we put these fields into dynamical theories of gravity to get a glimpse of the implications of their new non-trivial gravitationally induced (self)-interactions. The interesting case of realizing the inflaton as the time component of a Lorentz valued $0$-form shows already in the `torsion-less' limit that new venues of exploration are possible within this framework. A more careful analysis of this model is beyond the scope of this paper and will be developed in future work \cite{alfaroriquelme}.

\section*{Acknowledgements}
The work of JA is partially supported by Fondecyt 1110378  and Anillo ACT 1102. SR acknowledges support from Becas Chile and Fulbright-Conicyt scholarships and Nicol\'as Zalaquett for useful conversations.

\end{document}